\documentclass[preprint,11pt]{elsarticle}

\usepackage{lineno,hyperref}
\usepackage{amsmath}
\usepackage{amssymb}
\usepackage{amsbsy}
\usepackage{amsfonts}
\usepackage{amsopn}
\usepackage{amstext}
\usepackage{esint}
\usepackage{amssymb}
\usepackage{amsfonts}
\usepackage{amsmath}
\usepackage[english]{babel}
\usepackage{color}
\usepackage[dvips]{epsfig}
\usepackage[dvips]{graphicx}
\usepackage{float}
\usepackage{units}
\usepackage{textcomp}
\usepackage{babel}
\usepackage[dvips]{graphicx}
\usepackage{subfig}
\usepackage{hyperref}
\usepackage{slashed}
\usepackage{lineno,hyperref}
\modulolinenumbers[5]

\journal{Journal of \LaTeX\ Templates}

\begin{document}

\begin{frontmatter}

\title{A field theory in Randers-Finsler spacetime}

\author[mycorrespondingauthor]{J. E. G. Silva}
\address{Universidade Federal do Cariri (UFCA)\\
Av. Tenente Raimundo Rocha, \\ Cidade Universit\'{a}ria, Juazeiro do Norte, Cear\'{a}, CEP 63048-080, Brasil}
%\ead[url]{euclides.silva@ufca.edu.br}


\cortext[mycorrespondingauthor]{euclides.silva@ufca.edu.br}

%\address[ufca]{\inst{1} Universidade Federal do Cariri(UFCA), Av. Tenente Raimundo Rocha, \\ Cidade Universit\'{a}ria, Juazeiro do Norte, Cear\'{a}, CEP 63048-080, Brasil}

\begin{abstract}
Finsler geometry is a natural arena to investigate the physics of spacetimes with local Lorentz violating. The directional dependence of the Finsler metric provides a way to encode the Lorentz violating effects into the geometric structure of spacetime. Here, a classical field theory is proposed in a special Finsler geometry, the so-called Randers-Finsler spacetime, where the Lorentz violation is produced by a background vector field. By promoting the Randers-Finsler metric to a differential operator, a Finsler-invariant action for the scalar, gauge and fermions are proposed. The theory contains nonlocal terms, as in the Very Special Relativity based theories. By expanding the Lagrangian, minimal and nonminimal Standard Model Extension terms arises, revealing a perturbative Lorentz violation. For a CPT-even term, the Carrol-Field-Jackiw and derivative extensions are obtained.
\end{abstract}


\end{frontmatter}

%\linenumbers

\section{Introduction}
At Planck scale, several theories including string theory \cite{KS}, noncommutative geometry \cite{noncommutativegeometry}, Horava gravity \cite{horava}, and Very Special Relativity \cite{vsr} suggest that some low-energy symmetries, as the CPT and Lorentz symmetry, may be violated. A theoretical framework to study the effects of reminiscent Lorentz-CPT breaking at intermediate scales is the so-called Standard Model Extension (SME) \cite{cptviolation}. Lorentz violating effects studies have been carry out in a broad class of phenomena, ranging from muons, neutrinos, photons, hydrogen atom, among other %\cite{smeapplications}.
For a comprehensive review of Lorentz-violation tests, see Ref.\cite{rmp}.

The violation of the local Lorentz violation on gravity requires a modification on the local geometry of spacetime. In the context of the SME, local Lorentz violating coefficients arise due to vacuum expectation values (VEV) of self-interacting tensor fields in a Einstein-Cartan theory \cite{kosteleckysmegravity}.
% One way to overcome this is assuming the Lorentz violating tensors are created by the vacuum expectation value of a vector field, as the bumblebee \cite{bumblebee} or the aether fields \cite{Jacobson:2000xp} undergone a spontaneous symmetry process \cite{kosteleckysmegravity}. Another possibility is through an anisotropic geometry, the so-called Finsler geometry.
Another approach suggests that the modification of the dispersion relation of particles, an ubiquitous feature of Lorentz violating theories, arises due to an anisotropic Finsler geometry.

In Finsler geometry, the spacetime interval is measured with a general function, called Finsler function \cite{chern}. The Riemann geometry is a special case for which the Finsler function is quadratic on the vector components. The modified dispersion relations (MDR) are the starting point to propose a local Lorentz violating, as the DSR \cite{dsrfinsler} and VSR \cite{Gibbons}, whose the curved extension is the so-called Bogoslovsky spacetime \cite{bogoslovsky}. In the SME set up, the point particle classical Lagrangians and mass shell can be described by a class of Finsler functions, called SME-based Finsler geometries \cite{Kosteleckyriemannfinsler}. Each Lorentz-violating coefficient provides a different SME-based Finsler structure, as the b-space \cite{Kosteleckyriemannfinsler,foster}, bipartite \cite{Kosteleckybipartite,euclides}  and other spacetimes \cite{ColladayMc,russell,marco,marco2}.

A special Finsler structure is given by the Rander spacetime, where the anisotropy is driven by a background vector \cite{randers}. The Randers interval has the usual quadratic Lorentz interval added with a perturbative linear projection of the 4-velocity into the background vector \cite{randers}. In the context of the SME, the Randers spacetime can be regarded as a classical description of a fermion subjected to a CPT-Odd Lorentz-violating coefficient \cite{Kosteleckyriemannfinsler}. The particle dynamics in Randers spacetime is analogous to the Lorentzian invariant dynamics under the influence of a background electromagnetic vector field $a^\mu$. The effects of the Rander spacetime anisotropy have been studied in cosmology \cite{randerscosmology,randerscosmology2} and astrophysics \cite{randersastrophysics}.

The directional dependence of the Finsler metric leads naturally to a field theory living in the tangent bundle as the base space \cite{vacaru,pfeifer}. In this work we propose an alternative approach, where the fields live only on the spacetime. Since the Finsler metric depends on the momentum in the MDR, by replacing the momentum by the covariant derivative, the Finsler metric turns into a differential operator acting on the fields. Assuming a minimal coupling prescription to the Finsler metric, nonlocal operators, as those found in VSR \cite{vsr} and in Bogoslovksy spacetime \cite{bogoslovsky} are found. The perturbative character of the Randers anisotropy allow us to expand the nonlocal operators what yields to Lorentz violating terms as found in the SME \cite{cptviolation} and in the Carroll-Field-Jackiw model  higher derivative terms as found in the nonminimal SME \cite{gaugesmenonminimal,fermionsmenonminimal}.

This work is organized as the following. In section \ref{section1} the definition of the Randers spacetime and the dynamics of point particles are reviewed. In section \ref{section5} we propose e study the dynamics of scalar, gauge and fermion fields on the Randers spacetime. Final comments and perspectives are outlined in section \ref{section6}.

%%%%%%%%%%%%%%%%%%%%%%%%%%%%%%%%%%%%%%%%%%%%%%%%%%%%%%%%%%%%%%%%%%%%%%%%%%%%%%%%%%%%%%%%%%%%%%%%%%%%%%%%%%%%%%%%%%%%%%%%%%%%%%%%%%
%%%%%%%%%%%%%%%%%%%%%%%%%%%%%%%%%%%%%%%% END OF INTRODUCTION  %%%%%%%%%%%%%%%%%%%%%%%%%%%%%%%%%%%%%%%%%%%%%%%%%%%%%%%%%%%%%%%%%%%%
%%%%%%%%%%%%%%%%%%%%%%%%%%%%%%%%%%%%%%%%%%%%%%%%%%%%%%%%%%%%%%%%%%%%%%%%%%%%%%%%%%%%%%%%%%%%%%%%%%%%%%%%%%%%%%%%%%%%%%%%%%%%%%%%%%

\section{Randers-Finsler spacetime}
\label{section1}

%The Randers-Finsler geometry is a model to analyse the properties of the Lorentz violation by including a background vector field $a^\mu$ into the geometric structure of the spacetime.
In this section we review the definition and the main properties of the Randers-Finsler spacetime. We show how the background vector is included into the geometric structure of the spacetime and this yields to modification of the particle dynamics and mass shell.

In Randers-Finsler spacetime, the interval of two events $x^\mu,x^\mu +\dot{x}^\mu dt$ are measured by the so-called Finsler function,  $ds_R:=F_R(x,\dot{x})dt$, where \footnote{We adopt the mostly plus convention $(-,+,+,+)$ for the metric signature.} \cite{Kosteleckyriemannfinsler,randers}
\begin{eqnarray}
\label{randersinterval}
ds_{R}	&	:=	&	\left(\sqrt{-g_{\mu\nu}(x)\dot{x}^{\mu}\dot{x}^{\nu}} + \zeta a_{\mu}(x)\dot{x}^{\mu}\right)dt.
\end{eqnarray}
The Randers spacetime has the local Lorentz invariant interval $\alpha(x,\dot{x}):=\sqrt{-g_{\mu\nu}(x)\dot{x}^{\mu}\dot{x}^{\nu}}$ and a linear term $\beta(x,\dot{x}):=\zeta a_{\mu}(x)\dot{x}^{\mu}$ which 
drives the Lorentz violation. Both $g_{\mu\nu}(x)$ and $a_\mu (x)$ are considered background geometric tensors defining the metric properties of the anisotropic spacetime. The background Randers
vector $a$ has its indexes raised and lowed with the background Lorentzian metric $g_{\mu\nu}$.

The modified interval \ref{randersinterval} can be rewritten in terms of a anisotropic or Finsler metric, 
by $ds_{R}	=	\sqrt{-g_{\mu\nu}^F(x,\dot{x})\dot{x}^{\mu}\dot{x}^{\nu}}dt$ \cite{chern}, where the relation between the Finsler function $F(x,\dot{x})$ and the Finsler metric is given by
\begin{equation}
 g_{\mu\nu}^F(x,\dot{x})=-\frac{\partial F^2(x,\dot{x})}{\partial \dot{x}^{\mu} \partial \dot{x}^{\nu}}.
\end{equation}
In the Randers-Finsler spacetime, the Finsler metric has the form \cite{chern}
\begin{equation}
\label{randersmetric}
g_{\mu\nu}^{F}(x,\dot{x})=\frac{F}{\alpha}g_{\mu\nu} - \frac{\beta}{\alpha}u_{\mu}u_{\nu} + 2\zeta  u_{(\mu}a_{\nu)} + \zeta^{2}a_{\mu}a_{\nu},
\end{equation}
where $u_{\mu}:=g_{\mu\rho}u^\rho$ and $u^\rho:=\frac{\dot{x}^\rho}{\alpha(\dot{x})}$ is the Lorentzian unit 4-velocity.

The Randers-Finsler Lagrangian for a massive point particle is given by $L_{R}:=\left(-m\sqrt{-g_{\mu\nu}\dot{x}^{\mu}\dot{x}^{\nu}} + \zeta a_{\mu}(x)\dot{x}^{\mu}\right)$\cite{randers}, from which the 
canonical momentum covector has the form \cite{Kosteleckyriemannfinsler,changmdr}
\begin{equation}
\label{randersmomentumcovariant}
P_{\mu}^{R}	:=	\frac{\partial L^{R}}{\partial \dot{x}^{\mu}}=P_{\mu} - m\zeta a_{\mu},
\end{equation}
where $P_{\mu}:=mg_{\mu\nu}U^{\nu}$. The contravariant Randers-Finsler canonical momentum $P^R_\mu$ is given by $P^R_\mu:=g^F_{\mu\nu}(x,\dot{x})P^{R\nu}=P^{R\mu}:=m\frac{\dot{x}^{\mu}}{F(x,\dot{x})}$.

The modified mass shell (MDR) in the Finsler spacetimes is taken using the Finsler metric, i.e., \cite{bogoslovsky,girelli,vacaru,pfeifer}
\begin{equation}
\label{mdr}
g^{F}_{\mu\nu}(x,P^F)P^{F\mu}P^{F\nu}=-m^2.
\end{equation}
In the Randers spacetime, the MDR has the form $(g_{\mu\nu}+\zeta^2 a_{\mu}a_{\nu})P^{R\mu}P^{R\nu} - 2\zeta m a_{\mu}P^{R\mu}=-m^2$.
%\begin{equation}
%\label{randersmassshell}
%(g_{\mu\nu}+\zeta^2 a_{\mu}a_{\nu})P^{R\mu}P^{R\nu} - 2\zeta m a_{\mu}P^{R\mu}=-m^2.
%\end{equation}
The symmetries of the Finsler metric leads to the symmetries of the particles in this local Lorentz violating spacetime.
The (MDR)in Eq.\ref{mdr} is invariant by observer local Lorentz transformations and by deformed Lorentz transformations build with the Finsler Killing vectors. 

In a geodesic motion, the particle equation of motion (eom) is given by $\ddot{x}^{\mu}+G^\mu (x,\dot{x})=0$, where the inertial force is defined by $G^\mu = \gamma^{F\mu}_{\rho\sigma}(x,\dot{x})\dot{x}^{\rho}\dot{x}^\sigma$, and the Finsler Christoffel symbol is $\gamma^{F\mu}_{\nu\rho}(x,\dot{x}):=\frac{g^{F\mu\sigma}(x,\dot{x})}{2}\Big[\partial_{\nu}g^{F}_{\sigma\rho}(x,\dot{x})+\partial_{\rho}g^{F}_{\sigma\nu}(x,\dot{x})-\partial_{\sigma}g^{F}_{\nu\rho}(x,\dot{x})\Big]$. Let us seek for a covariant derivative for which the eom can be rewritten as $P^{R\mu}_{|\nu}=0$ and preserves the MDR \ref{mdr}.
Since the geometry, and hence the physics, is dependent on both the the position and velocities, consider the so-called horizontal derivative $\delta_{\mu}:=\frac{\partial}{\partial x^{\mu}} - N_{\mu}^{\nu}\frac{\partial}{\partial \dot{x}^\nu}$, and the vertical derivative $\bar{\partial
}_{\mu}:=F(x,\dot{x})\frac{\partial}{\partial \dot{x}^\mu}$, where $N^{\mu}_{\nu}:=\frac{\partial G^{\mu}}{\partial \dot{x}^\nu}$ \cite{chern}. Let us consider the Cartan connetion $\omega^{C\mu}_{\nu}:=\Gamma^{F\mu}_{\nu\sigma}dx^{\sigma} + C^{\mu}_{\nu\sigma}\delta y^{\mu}$, where $\Gamma^{F\mu}_{\nu\rho}(x,\dot{x}):=\frac{g^{F\mu\sigma}(x,\dot{x})}{2}\Big[\delta_{\nu}g^{F}_{\sigma\rho}(x,\dot{x})+\delta_{\rho}g^{F}_{\sigma\nu}(x,\dot{x})-\delta_{\sigma}g^{F}_{\nu\rho}(x,\dot{x})\Big]$ and $C_{\mu\nu\rho}(x,\dot{x}):=\frac{1}{2}\frac{\partial g^{F}_{\mu\nu}(x,\dot{x})}{\partial \dot{x}^{\rho}}$ is the so-called Cartan tensor, which measures the directional dependence of the Finsler metric \cite{chern}.
The Cartan horizontal covariant derivative of the Finsler metric tensor, defined as $g^{F}_{\mu\nu|\rho}:= \nabla^F_\rho g^{F}_{\mu\nu}= \delta_{\rho}g^{F}_{\mu\nu}-\Gamma^{F\sigma}_{\rho\mu}g^{F}_{\sigma\nu}-\Gamma^{F\sigma}_{\rho\nu}g^{F}_{\sigma\mu}$
vanishes, i.e., $g^{F}_{\mu\nu|\rho}\equiv 0$. Thus, a free particle experiences anisotropic local Lorentz violating inertial forces while preserving the MDR \ref{mdr}.

By applying the tetrad formalism, we can rewrite the Randers metric \ref{randersmetric} as $g_{\mu\nu}^F=E^{Fb}_\mu (x,\dot{x})E^{Fc}_\nu (x,\dot{x})\eta_{bc},
$ 
%\cite{chern}
%\begin{equation}\label{metricfactorization}
%g_{\mu\nu}^F=\Lambda^{Fb}_\mu (x,\dot{x})\Lambda^{Fc}_\nu (x,\dot{x})\eta_{bc},
%\end{equation}
where the tetrads are given by
\begin{eqnarray}
\label{randersvielbein}
E^{Fb}_\mu (x,\dot{x})&=&\sqrt{\frac{F}{\alpha}}\Big\{E^{a}_\mu + \left(\frac{\alpha}{F}\right)^2 \Big[-\frac{\beta}{2\alpha}U^\mu U_\mu +U^\mu a_b\nonumber\\
& +& a^\mu U_b + a^\mu a_b \Big] \Big\}\nonumber
\end{eqnarray}
The tetrad $E^{Fb}_\mu (x,\dot{x})$
can be understood as a deformation of the local Lorentz invariant tetrad $E^{b}_\mu (x,\dot{x})$ by the background field $a_\mu$.

\section{A Randers field theory}
\label{section5}

Once analysed the dynamics of a particle in Randers-Finsler spacetime, let us propose a dynamics for fields.
Likewise the particle action, we are interested in actions build from Randers tensors, as the Finsler metric. The relation $P_{\mu}^F=-i\delta_{\mu}$ suggests an approach where the direction dependence of the geometry becomes a momentum dependence of the metric. Assuming the fields have only position dependence, the momentum operator
has its origin on position variations, i.e., $P^F_{\mu}=-i \nabla^F_{\mu}$. Thus, the Finsler metric can be regarded as a differential operator, where $g^F_{\mu\nu}(x,y)\rightarrow g^F_{\mu\nu}(x,\nabla^F)$ and $y^\mu \rightarrow \nabla^{F\mu}$.
The relation $P^{F\mu}=\frac{\alpha}{F}P^\mu$ and the homogeneity of the Finsler metric, allow us to write the Finsler metric
as the operator $g^F_{\mu\nu}(x,\nabla)$. This approach of considering the Finsler metric as an operator is similar to the noncanonical kinetic terms \cite{noncanonicalkinetic}. Unlike the field theories defines on the tangent bunblde $TM_4$ \cite{pfeifer}, the Finsler metric operator procedure enable us to propose a field theory defined on the spacetime $M_4$ itself.

\subsection{Scalar field}

%A scalar field $\Phi(x)$ is naturally invariant over the Randers transformations, i.e., $\Phi(\Lambda_R x)=\Phi(x)$.
We propose a Randers invariant action for the scalar field is given by
\begin{eqnarray}
\label{finslerscalaraction}
S_{\Phi}:&=&-\frac{1}{2}\int_{M}\Big\{d^{4}x\{\nabla_{\mu}\Phi K^{\mu\nu}(x,\nabla) \nabla_{\nu}\Phi\nonumber\\
 &+& m^2 \sqrt{-g^F(x,\nabla)} \Phi^2]\}\Big\}.
\end{eqnarray}
where $K^{\mu\nu}:=\sqrt{-g^F(x,\nabla)}g^{F\mu\nu}(x,\nabla)$.
%\begin{equation}
%K^{\mu\nu}:=\sqrt{-g^F(x,\nabla)}g^{F\mu\nu}(x,\nabla).
%\end{equation}
%The scalar field action \ref{finslerscalaraction} is constructed with the Randers-Finsler scalars over the spacetime and not on the tangent bundle \cite{pfeifer}.
For $\zeta=0$, i.e., for a Local Lorentz symmetric spacetime, the action \ref{finslerscalaraction} yields to a minimal coupling of the scalar field in a curved spacetime.
% The position dependence of the background vector yields to the non-inertial effects of the Randers metric.

In the Randers spacetime, by means of the identification $y^\mu\rightarrow \nabla^\mu$, the contravariant metric has the form \cite{chern}
\begin{eqnarray}
g^{F\mu\nu}(x,\nabla)&=&\frac{ g^{\mu\nu}}{1+\zeta^2 a\cdot\nabla}-\frac{\zeta^2}{(1+\zeta^2 a\cdot\nabla)^2} (a^\mu \nabla^\nu\\ 
&+& a^\nu \nabla^\mu)+\frac{\zeta^4}{(1+\zeta^2 a\cdot\nabla)^3} \Big[a\cdot\nabla+ a^2\Big]\nabla^\mu \nabla^\nu \nonumber,
\end{eqnarray}
where $a\cdot\nabla:=a^\mu \nabla_\mu$. Then, defining a dimensionless background field $b^\mu:=\zeta a^\mu$, the Randers action for the scalar field can be rewritten as
\begin{equation}
\label{randersscalaraction1}
\begin{split}
S_{\Phi}&=-\frac{1}{2}\int_{M}d^{4}x \sqrt{-g}\Bigg\{ g^{\mu\nu}\nabla_\mu \Phi [(1+\zeta b\cdot\nabla)^{\frac{3}{2}}]\nabla_\nu \Phi\\
&+ (1+\zeta^2 a\cdot\nabla)^{\frac{5}{2}}m^2 \Phi^2 +
\zeta\nabla_\mu \Phi \Big[(1+\zeta b\cdot\nabla)^{\frac{3}{2}}(g^{\rho(\mu}b^\nu)\Big]\times\\
&\nabla_\rho \nabla_\nu \Phi +  \zeta^2 g^{\mu\rho}g^{\nu\sigma}\nabla_\mu \Phi \Big[\frac{(\zeta \beta + b^2)}{(1+\zeta b\cdot\nabla)^{\frac{1}{2}}}\Big]\nabla_\nu \nabla_\rho \nabla_\sigma \Phi \Bigg\} .
\end{split}
\end{equation}
%\begin{eqnarray}
%S_{\Phi}&=&-\frac{1}{2}\int_{M}d^{4}x \sqrt{-g}\Bigg\{g^{\mu\nu}\nabla_\mu \Phi [(1+\zeta b\cdot\nabla)^{\frac{3}{2}}]\nabla_\nu \Phi + (1+\zeta^2 a\cdot\nabla)^{\frac{5}{2}}m^2 \Phi^2 \\
%&+& \zeta\nabla_\mu \Phi \Big[(1+\zeta b\cdot\nabla)^{\frac{3}{2}}(g^{\mu\rho}b^\nu + g^{\nu\rho}b^\mu)\Big]\nabla_\rho \nabla_\nu \Phi + \zeta^2 g^{\mu\rho}g^{\nu\sigma}\nabla_\mu \Phi \Big[\frac{(\zeta \beta + b^2)}{(1+\zeta b\cdot\nabla)^{\frac{1}{2}}}\Big]\nabla_\nu \nabla_\rho \nabla_\sigma \Phi \Bigg\} \nonumber.
%\end{eqnarray}
The action exhibits nonlocal dynamical terms similar to those of VSR \cite{vsr}.
The perturbative character of the Randers spacetime allow us to rewrite the Randers Lagrangian as
\begin{equation}
\mathcal{L}^{F}_\phi=\mathcal{L}_{LI}+\zeta \mathcal{L}^{1}_{LV}+\zeta^2 \mathcal{L}^{2}_{LV} +  \cdots,
\end{equation}
where $\mathcal{L}_{LI}:=-\frac{\sqrt{-g}}{2}(g^{\mu\nu}\nabla_\mu \Phi\partial_\nu \Phi + m^2 \Phi)$ is the usual Lorentz-invariant Klein-Gordon Lagrangian. The first-order Lorentz-violating Lagrangian is given by
\begin{equation}
\label{scalarlvlagrangian}
\mathcal{L}^{1}_{LV}= \nabla_\mu \Phi (K^{5})^{\mu\nu}\nabla_\nu \Phi - \frac{5m^2 \sqrt{-g}}{4} b^\rho \nabla_\rho (\Phi^2),
\end{equation}
where the mass dimension five Lorentz violating operator $(K^{5})^{\mu\nu}$ has the form
\begin{equation}
(K^{5})^{\mu\nu}:=-\frac{\sqrt{-g}}{4}\Big\{3 g^{\mu\nu}b^\rho - 2 (g^{\mu\rho}b^\nu + g^{\nu\rho}b^\mu) \Big\}\nabla_\rho.
\end{equation}
The second-order terms forms the Lorentz-violating Lagrangian
\begin{eqnarray}
\label{scalarlvlagrangian2}
\mathcal{L}^{2}_{LV}= \nabla_\mu \Phi (K^{6})^{\mu\nu}\nabla_\nu \Phi - \frac{15m^2 \sqrt{-g}}{8} b^\rho b^\sigma \nabla_\rho\nabla_\sigma (\Phi^2)\nonumber,
\end{eqnarray}
where the mass dimension six operator $(K^{6})^{\mu\nu}$ is defined as
\begin{equation}
(K^{6})^{\mu\nu}:=-\frac{\sqrt{-g}}{4}\Big\{\frac{3}{8} g^{\mu\nu}b^\rho b^\sigma - \frac{b^2}{2} g^{\mu\rho}g^{\nu\sigma}\Big\}\nabla_\rho \nabla_\sigma.
\end{equation}
It is worthwhile to say that the last term in the first-order perturbed Lagrangian Eq.\ref{scalarlvlagrangian}, for a covariantly constant background vector $b^\mu$, provides a total derivative term which can be dropped from the action.
The operators $(K^{5})^{\mu\nu}$ and $(K^{6})^{\mu\nu}$ in a flat background metric have the same form of the nonminimal Standard Model Extension for dimension 5 and 6 Lorentz violating operators.

%%%%%%%%%%%%%%%%%%%%%%%%%%%%%%%%%%%%%%%%%%%%%%%%%%%%%%%%%%%%%%%%%%%%%%%

\subsection{Vector Gauge field}

Let us propose a Finslerian dynamics for the Abelian gauge field such that the field strength is defined using Cartan horizontal covariant derivatives instead of the Levi-Civita covariant derivative, i.e.,
\begin{equation}
F^{R}_{\mu\nu}:=\nabla^F_\mu A_\nu - \nabla^F_\nu A_\mu = A_{\nu|\mu}-A_{\mu|\nu},
\end{equation}
where
$\nabla^F_\mu A_\nu:=\delta_\mu A_\mu - \Gamma^{F\rho}_{\mu\nu}A_\rho$.
%Using the Cartan connection, the Finslerian field strength can be rewritten as $F^{R}_{\mu\nu}=\delta_{[\mu}A_{\nu]}$.
%Under a gauge transformation $A'_\mu=A_\mu + \delta_\mu f$, the field strength changes as $F'^{R}_{\mu\nu}=F^R_{\mu\nu}+[\delta_\mu , \delta_\nu]f$. 
Since in our approach, $A=A(x)$, then $\delta_\mu = \partial_\mu$ and $F^{R}_{\mu\nu}=F_{\mu\nu}$. Therefore, the electric and magnetic components of the field strength and the gauge symmetry are preserved in Randers spacetime.

A gauge Finslerian extension of the Maxwell action has the form
\begin{eqnarray}
\label{randersgaugeaction}
S^F_{A}&:=&-\frac{1}{4}\int d^4 x\Big\{F_{\mu\nu}K^{F\mu\nu\rho\sigma}(x,\nabla)F_{\rho\sigma}\Big\},
\end{eqnarray}
where $K^{F\mu\nu\rho\sigma}:=\sqrt{-g^F(x,\nabla)}g^{F\mu\rho}(x,\nabla)g^{F\nu\sigma}(x,\nabla)$.
For $\zeta=0$, the tensor $K^{F\mu\nu\rho\sigma}$ turns into $\sqrt{-g(x)}g^{\mu\rho}(x)g^{\nu\sigma}(x)$ we obtain the usual Maxwell term $\mathcal{L}_A=-\frac{\sqrt{-g(x)}}{4}g^{\mu\rho}(x)g^{\nu\sigma}(x)F_{\mu\nu}F_{\rho\sigma}$.

In Randers spacetime, the Finsler gauge action \ref{randersgaugeaction} yields to the Finsler gauge Lagrangian
\begin{eqnarray}
\label{randersgaugelagrangian}
\mathcal{L}^{F}_{A}&=&-\frac{\sqrt{-g}}{4}F_{\mu\nu}\Big\{(1+\zeta b\cdot\nabla)^{\frac{1}{2}}g^{\mu\rho}g^{\nu\sigma}\nonumber\\ 
&-&2\zeta^2 \Big[\frac{g^{\mu\rho}g^{\nu\lambda}g^{\sigma\xi}}{(1+\zeta b\cdot \nabla)^{\frac{1}{2}}}\Big]\nonumber\\
&+& 2\zeta^2 \Big[\frac{g^{\mu\rho}g^{\nu\lambda}g^{\sigma\xi}(\zeta \beta + b^2)}{(1+\zeta b\cdot \nabla)^{\frac{3}{2}}}\nabla_\lambda \nabla_\xi \Big]\Big\}F_{\rho\sigma}.
\end{eqnarray}
%The Finsler gauge action in Eq.\ref{randersgaugelagrangian} exhibits nonlocal operators.
Let us analyse the gauge Lagrangian
in Eq.\ref{randersgaugelagrangian} term by term.
The first term
\begin{equation}
\mathcal{L}^1_{A}=-\frac{\sqrt{-g}}{4}F_{\mu\nu}(1+\zeta b\cdot\nabla)^{\frac{1}{2}}g^{\mu\rho}g^{\nu\sigma}F_{\rho\sigma},
\end{equation}
can be expanded in powers of $\zeta$ as
\begin{equation}
\mathcal{L}^1_{A}=-\frac{\sqrt{-g}}{4}F^{\mu\nu}F_{\mu\nu} + \zeta F_{\mu\nu}(\hat{k}^{(5)}_F)^{\mu\nu\rho\sigma}F_{\rho\sigma}+ \cdots ,
\end{equation}
where the zero order term consists of the usual Maxwell Lagrangian and the first-order and second-order Lorentz-violating Lagrangian $\mathcal{L}_{ALV}$ term are respectively, $(\hat{k}^{(5)}_F)^{\mu\nu\rho\sigma}:=-\frac{\sqrt{-g}}{8}g^{\mu\rho}g^{\nu\sigma}b^\lambda \nabla_\lambda$ and $(\hat{k}^{(6)}_F)^{\mu\nu\rho\sigma}:=-\frac{\sqrt{-g}}{8}g^{\mu\rho}g^{\nu\sigma}b^\lambda b^\xi \nabla_\lambda \nabla_\xi$ \cite{gaugesmenonminimal}.
The second term
\begin{equation}
\mathcal{L}^2_{A}:=\frac{\sqrt{-g}}{2}\zeta^2 F_{\mu\nu}\Bigg[\frac{g^{\mu\rho}g^{\nu\lambda}g^{\sigma\xi}}{(1+\zeta b\cdot \nabla)^{\frac{1}{2}}}(\nabla_{(\lambda} b_{\xi)})\Bigg]F_{\rho\sigma},
\end{equation}
can be expanded as
\begin{equation}
\mathcal{L}^2_{A}=\zeta^2 [F_{\mu\nu}(\hat{k}_F^{5})^{\mu\nu\rho\sigma}F_{\rho\sigma}+F_{\mu\nu}\tilde{k}^{\mu\nu\rho\sigma}F_{\rho\sigma}]+\cdots,
\end{equation}
%\zeta^3 [F_{\mu\nu}(k^{6})^{\mu\nu\rho\sigma}F_{\rho\sigma}+F_{\mu\nu}\tilde{\tilde{k}}^{\mu\nu\rho\sigma}F_{\rho\sigma}] +
where $(\hat{k}_F^{5})^{\mu\nu\rho\sigma}:=\frac{\sqrt{-g}}{2}g^{\mu\rho}(g^{\nu\lambda}b^\sigma  + g^{\sigma\lambda}b^\nu)\nabla_\lambda$ and $\tilde{k}^{\mu\nu\rho\sigma}:=\frac{\sqrt{-g}}{2}g^{\mu\rho}[(\nabla^\nu b^\sigma + \nabla^\sigma b^\rho)]$. It is worth to note that $\tilde{k}^{\mu\nu\rho\sigma}=0$ for a covariantly constant background vector, i.e., for a Randers spacetime of Berwald type. For a background flat spacetime and constant background vector, the mass dimension five operator
$(\hat{k}_F^{5})^{\mu\nu\rho\sigma}$ belongs to the nonminimal SME \cite{gaugesmenonminimal}.

The third term
\begin{equation}
\mathcal{L}^2_{A}:=\frac{\sqrt{-g}}{2}\zeta^2 F_{\mu\nu}\Bigg[\frac{g^{\mu\rho}g^{\nu\lambda}g^{\sigma\xi}(\zeta \beta + b^2)}{(1+\zeta b\cdot \nabla)^{\frac{3}{2}}}\nabla_\lambda \nabla_\xi \Bigg]F_{\rho\sigma},
\end{equation}
can be rewritten as
\begin{equation}
\mathcal{L}^3_{A}:=\zeta^2 F_{\mu\nu}(\hat{k}_F^{(6)})^{\mu\nu\rho\sigma}F_{\rho\sigma},
\end{equation}
where $(\hat{k}_F^{(6)})^{\mu\nu\rho\sigma}:=-\frac{\sqrt{-g}}{2}b^2 g^{\mu\rho}g^{\nu\lambda}g^{\sigma\xi}\nabla_\lambda \nabla_\xi$,
for a flat background metric and vector, is a dimension six Lorentz coefficient of the nonminimal SME \cite{gaugesmenonminimal}.

%Another gauge and Finsler invariant Lagrangian is given by
%\begin{equation}
%\label{finslergaugeaction2}
%\tilde{S}^F:=\int d^4 x[\sqrt{-g^F(x,\nabla)}\epsilon^{\mu\nu\rho\sigma}F_{\mu\nu}F_{\rho\sigma}],
%\end{equation}
%that in Randers spacetime yields to $\mathcal{\tilde{L}}^F=(1+\zeta^2 a\cdot\nabla)^{\frac{3}{2}}\epsilon^{\mu\nu\rho\sigma}%F_{\mu\nu}F_{\rho\sigma}$. Expanding this Finsler gauge Lagrangian, we obtain
%\begin{equation}
%\mathcal{\tilde{L}}^F=\mathcal{\tilde{L}}_{LI}+\Big[\zeta (\hat{k}^{(5)}_F)^{\mu\nu\rho\sigma}+\zeta^2 (\hat{k}^{(6)}_F)^{\mu\nu%\rho\sigma} + \cdots \Big]F_{\mu\nu}F_{\rho\sigma},
%\end{equation}
%where $\mathcal{\tilde{L}}_{LI}=\epsilon^{\mu\nu\rho\sigma}F_{\mu\nu}F_{\rho\sigma}$ , $(\hat{k}^{(5)}_F)^{\mu\nu\rho\sigma}:=\frac{3}{2}\epsilon^{\mu\nu\rho\sigma}b^\lambda \nabla_\lambda$ and $(\hat{k}^{(6)}_F)^{\mu\nu\rho\sigma}:=-\frac{3}{4}\epsilon^{\mu\nu\rho\sigma}b^\lambda b^\epsilon \nabla_\lambda \nabla_\epsilon$.

The Randers geometry also allows the following gauge invariant action coupling 
\begin{equation}
\label{finslergaugeaction3}
\hat{S}^F:=\zeta \int d^4 x[\epsilon^{\mu\nu\rho\sigma}A_\mu \hat{K}_\nu F_{\rho\sigma}],
\end{equation}
where $\hat{K}_\nu := \sqrt{-g^F(x,\nabla)} a_\nu$. Expanding the action in Eq.\ref{finslergaugeaction3}, we obtain the Lorentz violating Lagrangian
\begin{equation}
\begin{split}
\hat{\mathcal{L}}^F=&\zeta \epsilon^{\mu\nu\rho\sigma}A_\mu a_\nu F_{\rho\sigma} + \zeta^2 \epsilon^{\mu\nu\rho\sigma}a^\lambda A_\mu \nabla_\lambda a_\nu F_{\rho\sigma}\\
&+\zeta^2 \epsilon^{\mu\nu\rho\sigma}a^\lambda A_\mu  a_\nu \nabla_\lambda F_{\rho\sigma} + \cdots,
\end{split}
\end{equation}
i.e., the so-called Carrol-Field-Jackiw Lagrangian \cite{Carroll} and corrections.

%Therefore, the Randers actions Eq.\ref{randersgaugeaction} and Eq.\ref{finslergaugeaction2} for the vector gauge field can be regarded as a series of Lorentz violating terms of the nonminimal Standard Model Extension \cite{gaugesmenonminimal}.

%%%%%%%%%%%%%%%%%%%%%%%%%%%%%%%%%%%%%%%%%%%%%%%%%%%%%%%%%%%%%%%%%%%%%%%%%%%%%%%%%%%%%%%%%%%%%%%%%%%%%%%%%%%%%%%%%%%%%%%%%%%%%%%%%%%%%%%%%%%%%%%%%%%%%%%%%%%%%%%%%%%%%%%%%%%%%%%5
%%%%%%%%%%%%%%%%%%%%%%%%%%%%%%%%%%%%%%%%%%%%%%%%%%%%%%%%%%%%%%%%%%%%%%%%%%%%%%%%%%%%%%%%%%%%%%%%%%%%%%%%%%%%%%%%%%%%%%%%%%%%%%%%%%%%%%%%%%%%%%%

\subsection{Fermionic field}

In order to describe fermions, we adopt the tetrad formalism, whereby the Randers gamma matrices are defined as $\gamma^{F\mu}(x,P):=E^{F\mu}_{b}(x,P)\gamma^b$ and 
the tetrads are defined in Eq.\ref{randersvielbein}. 
As done for the Finsler metric, the Finsler gamma matrices are seen as operators, defined as
\begin{equation}
\hat{\gamma}^{F\mu}(D):=\sqrt{-g^{F}(D)}\gamma^{F\mu}(D).
\end{equation}
A Finsler-invariant Dirac action has the form
\begin{equation}
\label{finslerdiracaction}
S^F_{\Psi}:=\int_{M_4} d^4x\Big[\bar{\Psi}[i\hat{\gamma}^{F\mu}(D)D_\mu -m]\Psi + H.C. \Big],
\end{equation}
where the Finsler covariant derivative $D_\mu$ is defined as
\begin{equation}
D^F_\mu:=\partial_\mu +\frac{i}{2}\Gamma^{Fbc}_{\mu}\Sigma_{bc},
\end{equation}
and the local-flat generators $\Sigma_{bc}$ are defined as $\Sigma_{bc}:=\frac{i}{4}[\gamma_b , \gamma_c]$.
For a constant background metric $\eta_{\mu\nu}$, the Finslerian connection coefficients depends only on the derivatives of the background vector $b_\mu=\zeta a_\mu$. Therefore, assuming a constant background vector, we adopt $D_\mu = \partial_\mu$.
In Randers spacetime the Finsler Lagrangian for the fermion takes the form
\begin{eqnarray}
\label{randersfermionlagrangian}
\mathcal{L}^F_{\Psi}&=&\bar{\Psi}(1+\zeta b\cdot \partial)^{\frac{5}{2}}\Bigg\{\frac{i}{(1+\zeta b\cdot \partial)^{\frac{1}{2}}}\Big[\gamma^\mu \partial_\mu \nonumber\\
&-& \frac{\zeta}{2}\frac{(\gamma^\mu b^\nu \partial_\mu \partial_\nu + \gamma^\mu \eta^{\nu\rho}b_\mu \partial_\nu \partial_\rho)}{(1+\zeta b\cdot \partial)}+\\
 &+&\frac{\zeta^2}{2}\frac{(\zeta b\cdot \partial +b^2)}{(1+\zeta b\cdot \partial)^2}\gamma^\mu \eta^{\nu\rho}\partial_\mu \partial_\nu\partial_\rho \Big] -m\Bigg\}\Psi + H.C. \nonumber.
\end{eqnarray}
It is worthwhile to note the presence of nonlocal operators as in VSR \cite{vsr} and in Bogoslovksky spacetime.

Expanding the nonlocal operators we can rewrite Finsler-Dirac Lagrangian \ref{randersfermionlagrangian} as
\begin{equation}
\mathcal{L}^F_{\Psi}=\bar{\Psi}(i \gamma^\mu \partial_\mu - m + \hat{\mathcal{Q}})\Psi,
\end{equation}
where $\hat{\mathcal{Q}}$ is a Lorentz violating operator. $\hat{\mathcal{Q}}$ can be written as
\begin{equation}
\hat{\mathcal{Q}}=\zeta \Big[\mathcal{Q}^{(4)} + \hat{a}^{(5)\mu}\gamma_\mu\Big] + \cdots,
\end{equation}
where $\mathcal{Q}^{(4)}$ is a mass dimension four term 
\begin{equation}
\mathcal{Q}^{(4)}:=-\frac{5}{2}m b^\mu \partial_\mu,
\end{equation}
and $\hat{a}^{(5)\mu}$ is CPT-Odd vector operators of mass dimension five given by
\begin{equation}
\hat{a}^{(5)\mu}:=(b^\nu \partial^\mu \partial_\nu - \eta^{\nu\rho}b^\mu \partial_\nu \partial_\rho)
\end{equation}
%\begin{eqnarray}
%\hat{a}^{(7)\mu}:&=& \zeta^3\Big[\frac{5}{4}b^\nu b^\rho b^\sigma \partial^\mu \partial_\nu\partial_\rho\partial_\sigma - b_\mu b^\sigma b^\lambda \Box\partial_\sigma\partial_\lambda\Big] ,
%\zeta Q^{(2)} + \zeta^2 Q^{(3)} + \cdots\\
%\end{eqnarray}
%$\hat{c}^{(6)\mu}$ is CPT-even mass dimension six vector given by
%\begin{equation}
%\hat{c}^{(6)\mu}:=\Big[\frac{1}{2}\partial^\mu b^\nu b^\rho \partial_\nu \partial_\rho - \frac{1}{2} b^\mu \eta^{\nu\rho}b^\sigma \partial_\nu \partial_\rho\partial_\sigma + b^2 \partial^\mu \eta^{\nu\rho}\partial_\nu \partial_\rho\Big],
%\end{equation}
%and $\hat{e}^{(6)}$ is scalar CPT-Odd of mass dimension six defined as
%\begin{eqnarray}
%\hat{e}^{(6)}&:=&-  \frac{15m}{4} b^\mu b^\nu b^\rho \partial_\mu \partial_\nu   \partial_\rho.
%\end{eqnarray}
%\begin{eqnarray}
%\hat{e}:= \zeta^3 \frac{15}{8}m b^\mu b^\nu b^\rho \partial_\mu \partial_\nu\partial_\rho
%\end{eqnarray}
Therefore, the Finsler-Dirac action in Eq.\ref{finslerdiracaction} yields to Lorentz violating terms similar to those
of the nonminimal SME \cite{fermionsmenonminimal}. 

%The structure of the Finsler-Dirac Lagrangian \ref{randersfermionlagrangian} suggests the only
%nonminimal SME terms arising in the Randers spacetime are the scalars and vectors.

%%%%%%%%%%%%%%%%%%%%%%%%%%%%%%%%%%%%%%%%%%%%%%%%%%%%%%%%%%%%%%%%%%%%%%%%%%%%%%%%%%%%%%%%%%%%%%%%%%%%%%%%%%%%%%%%%%%%%%%%%%%%%%%%%%%%%%%%%%%%%%%%%%%
%%%%%%%%%%%%%%%%%%%%%%%%%%%%%%%%%%%%%%%%%%%%%%%%%%%%%%%%%%%%%%%%%%%%%%%%%%%%%%%%%%%%%%%%%%%%%%%%%%%%%%%%%%%%%%%%%%%%%%%%%%%%%%%%%%%%%%%%%%%%%%%%%
\section{Final remarks and Perspectives}
\label{section6}

We defined a Finsler field theory based on the Randers spacetime. By interpreting the Finsler metric as a differential operator acting on the fields, we showed that a minimal coupling of the scalar and vector fields with the Randers-Finsler metric, and the coupling of the fermion with the Randers-Finsler vielbein yields to nonlocal operators. The presence of noncanonical or nonlocal terms in this coupling is similar to those found in VSR \cite{vsr} and in the Bogoslovsky spacetime \cite{bogoslovsky}.  The perturbative character of the Lorentz violating allow us to expand these nonlocal operators and find some minimal and nonminiamal SME terms. For the gauge vector field, the Randers background vector also produces the Carroll-Field-Jackiw term and corrections.  As perspectives we point out the analysis of bounds for the coupling proposed based on tests for the gauge \cite{gaugesmenonminimal}, fermionic \cite{fermionsmenonminimal}. The stability analysis of the fields and the gravitational sector are also important future developments.

%\acknowledgments
\section*{Acknowledgments}
\hspace{0.5cm}The author thank the Conselho Nacional de Desenvolvimento Cient\'{\i}fico e Tecnol\'{o}gico (CNPq), grants n$\textsuperscript{\underline{\scriptsize o}}$ 312356/2017-0

\end{document}